\documentclass[aps,prd,twocolumn,superscriptaddress,amsmath]{revtex4-1}

\begin{document}

\title{Rotating (A)dS black holes in bigravity}

\author{Eloy Ay\'on-Beato}
\email{ayon-beato-at-fis.cinvestav.mx} \affiliation{Departamento de
F\'{\i}sica, CINVESTAV--IPN, Apdo. Postal 14--740, 07000 M\'exico~D.F.,
M\'exico} \affiliation{Instituto de Ciencias F\'{\i}sicas y Matem\'aticas,
Universidad Austral de Chile, Casilla 567 Valdivia, Chile}

\author{Daniel Higuita-Borja}
\email{dhiguita-at-fis.cinvestav.mx} \affiliation{Departamento de
F\'{\i}sica, CINVESTAV--IPN, Apdo. Postal 14--740, 07000 M\'exico~D.F.,
M\'exico} \affiliation{Instituto de Ciencias F\'{\i}sicas y Matem\'aticas,
Universidad Austral de Chile, Casilla 567 Valdivia, Chile}

\author{Julio A. M\'endez-Zavaleta}
\email{jmendezz-at-fis.cinvestav.mx} \affiliation{Departamento de
F\'{\i}sica, CINVESTAV--IPN, Apdo. Postal 14--740, 07000 M\'exico~D.F.,
M\'exico} \affiliation{Instituto de Ciencias F\'{\i}sicas y Matem\'aticas,
Universidad Austral de Chile, Casilla 567 Valdivia, Chile}

\begin{abstract}
In this paper we explore the advantage of using the Kerr-Schild \emph{Ansatz} in
the search of analytic configurations to bigravity. It turns out that it
plays a crucial role by providing means to straightforwardly calculate the
square root matrix encoding the interaction terms between both gravities.
We rederive in this spirit the Babichev-Fabbri family of asymptotically
flat rotating black holes with the aid of an emerging circularity theorem.
Taking into account that the interaction terms contain by default two
cosmological constants, we repeat our approach starting from the more
natural seeds for the Kerr-Schild \emph{Ansatz} in this context: the (A)dS
spacetimes. As result, we show that a couple of Kerr-(A)dS black holes
constitute an exact solution to ghost free bigravity. These black holes
share the same angular momentum and (A)dS radius but their masses are not
constrained to be equal, similarly to the asymptotically flat case.
\end{abstract}

\maketitle

%%%%%%%%%%%%%%%%%%%%%%%%%%%%%%%%%%%%%%%%%%%%%%%%%%%%%%%%%%%%%
\section{Introduction\label{Sec:Intro}}
%%%%%%%%%%%%%%%%%%%%%%%%%%%%%%%%%%%%%%%%%%%%%%%%%%%%%%%%%%%%%

Almost five years ago de Rham, Gabadadze and Tolley proposed a ghost-free and
consistent interaction potential for massive gravity \cite{deRham:2010kj}
giving rise to what is now known as the dRGT theory. One of the distinctive characteristics of their construction of massive gravity is the need for a reference metric, due to the impossibility of constructing nonderivative self-interactions with
the dynamical metric only. Later, Hassan and Rosen \cite{Hassan:2011zd}
showed that this arbitrary metric can be promoted to be dynamical too and the
resulting theory would still be free of the Boulware-Deser ghost
\cite{Boulware:1973my}, propagating a total of $5+2$ degrees of freedom; this
theory became what is now called bigravity.

Thanks to overcoming the theoretical difficulties manifest in theories
including massive gravitons (see e.g.\
\cite{Hinterbichler:2011tt,deRham:2014zqa} for complete reviews on the
subject), the interest of the community in these models has increased in recent years   resulting in numerous studies on a wide variety of
topics. For example, effort has been made to construct cosmological models
that can explain the accelerated expansion of the Universe as a natural
consequence of regarding the interaction of gravity through a massive boson, although no viable and stable cosmological solutions have been reported so far
\cite{deRham:2011by, Volkov:2011an,vonStrauss:2011mq,Volkov:2012cf,Akrami:2012vf,Konnig:2013gxa}.
Another important branch of interest lies in finding exact configurations
supporting the massive generalization of the gravitational equations. In
this category black holes solutions \cite{Koyama:2011xz,Comelli:2011wq,%
Nieuwenhuizen:2011sq,Berezhiani:2011mt,Volkov:2012wp,Brito:2013xaa,%
Babichev:2014fka, Enander:2015kda,Tolley:2015ywa} play an important role since due to the new degrees of
freedom their stability properties are notably different from those
characterizing the final state of the gravitational collapse in general
relativity. See Ref.~\cite{Babichev:2015xha} for a complete review of the
plethora of such solutions to massive (bi)gravity.

Recently the first rotating black hole solutions in massive gravity were
presented by Babichev and Fabbri \cite{Babichev:2014tfa}; they showed the
Kerr spacetime \cite{Kerr:1963ud} is a solution to the dRGT theory with a flat
reference metric if a precise relation between the coupling constants of the
theory holds. This result is also true for bigravity. In this case, the
solution consists of two copies of the Kerr black hole with the same angular
momenta but not necessarily equal masses. It is also possible to charge the
solution through a coupling between the electric field to only one of the
metrics without adding undesirable degrees of freedom.

The finding by Babichev and Fabbri of rotating configurations in bigravity  is
a major step in the comprehension of the dynamics of this theory due to the
great difficulty involved in the calculation of the ghost-free interaction
terms between the involved metrics. However, the fact that the interaction
terms naturally include cosmological constants for each metric makes it more
expected that the resulting configurations be asymptotically (Anti-)de Sitter
[(A)dS] instead of the asymptotically flat behavior unveiled by Babichev and
Fabbri. In the present work, we aim to build this generalization by deriving
a class of asymptotically (A)dS rotating black holes which are analogue to
that originally found by Carter in \cite{Carter:1968ks}. We will take
advantage of the generalized Kerr-Schild \emph{Ansatz} to integrate the field
equations of bigravity, making transparent how both Kerr-(A)dS black holes
appear.

In Sec.~\ref{Sec:Bigravity} we briefly introduce our framework which
corresponds to the Hassan-Rosen bigravity \cite{Hassan:2011zd} with the
coupling parameters left free up to the Fierz-Pauli limit. Next, the
asymptotically flat solution of Babichev and Fabbri \cite{Babichev:2014tfa}
will be rederived to illustrate our procedure. We start in
Sec.~\ref{Sec:Kerr-Schild} by considering one metric as a Kerr-Schild
transformation and the other as proportional to a Kerr-Schild transformation,
both starting from Minkowski spacetime and possessing different profile
functions. We show how easy it is to calculate the interaction terms with the
help of this \emph{Ansatz}. As the direct integration of the involved profiles is
far from straightforward, a geometrical approach proving the circularity of
these stationary axisymmetric configurations is developed in
Sec.~\ref{sec:Circularity}, which will result extremely useful by fixing the
angular dependence of the profiles and making straightforward the integration
of the remaining bigravity equations. In Sec.~\ref{Sec:Kerr-deSitter} we will
follow an analogue procedure but replace the Minkowski spacetime with the (A)dS
one as the starting seed of the generalized Kerr-Schild transformation. Using
similar circularity arguments, we prove that two Kerr-(Anti-)de Sitter black
holes with same (A)dS radii and angular momenta but arbitrary masses are
solutions of bigravity up to three constraints in the coupling constants of
the theory, which generalize the constraints previously found by Babichev and
Fabbri. A discussion of the results and the difficulties involved in their
generalization is included in  Sec.~\ref{sec:discussion}.

%%%%%%%%%%%%%%%%%%%%%%%%%%%%%%%%%%%%%%%%%%%%%%%%%%%%%%%%%%%%%
\section{Ghost-free bigravity\label{Sec:Bigravity}}
%%%%%%%%%%%%%%%%%%%%%%%%%%%%%%%%%%%%%%%%%%%%%%%%%%%%%%%%%%%%%

Bigravity as formulated by Hassan and Rosen \cite{Hassan:2011zd} is a
four-dimensional ghost-free theory describing two metric fields $g_{\mu\nu}$
and $f_{\mu\nu}$ interacting via a nonderivative potential. One of these
fields is massive and the other is not; hence the theory propagates a total
of $5+2$ degrees of freedom. The interaction is encoded through scalar
quantities computed from a matrix defined by the following quadratic relation
\begin{equation}\label{def:gamma}
{(\gamma^2)^\mu}_\nu={\gamma^\mu}_\alpha{\gamma^\alpha}_\nu
\equiv g^{\mu\alpha}f_{\alpha\nu}.
\end{equation}
The bigravity defining action is
\begin{subequations}\label{eq:action}
\begin{align}
S[g,f]={}& \frac{1}{2\kappa_g}\int d^4x\sqrt{-g}R[g]+\frac{1}{2\kappa_f}\int
d^4x\sqrt{-f}\mathcal{R}[f]\nonumber\\
        & -\frac{m^2}{\kappa}\int d^4x\sqrt{-g}\mathcal{U}[g,f],
   \label{eq:Bigravity}
\end{align}
where $R[g]$ and $\mathcal{R}[f]$ are the Ricci scalars for each metric,
$\kappa_{g}$ and $\kappa_{f}$ are the corresponding Einstein constants,
$\kappa$ is a function of $\kappa_g$ and $\kappa_f$ with the same dimensions,
and $m$ is the graviton mass. The interaction between the metrics is
mediated by the potential
\begin{equation}\label{eq:Int}
\mathcal{U}[g,f]=\sum_{k=0}^{4}b_k\mathcal{U}_k(\gamma),
\end{equation}
where $b_k$ are coupling constants and the interaction terms are defined by
\begin{align}
\mathcal{U}_0(\gamma)&=1, \qquad
\mathcal{U}_1(\gamma) =\sum_A\lambda_A=\left[\gamma\right],\nonumber\\
\mathcal{U}_2(\gamma)&=\sum_{A<B}\lambda_A\lambda_B
                      =\frac{1}{2!}\left([\gamma]^2-[\gamma^2]\right),
                       \nonumber\\
\mathcal{U}_3(\gamma)&=\sum_{A<B<C}\lambda_A\lambda_B\lambda_C
                      =\frac{1}{3!}\left([\gamma]^3-3[\gamma][\gamma^2]
                       +2[\gamma^3]\right),\nonumber\\
\mathcal{U}_4(\gamma)&=\lambda_0\lambda_1\lambda_2\lambda_3
                      =\frac{1}{4!}\left([\gamma]^4-6[\gamma]^2[\gamma^2]
                       +8[\gamma][\gamma^3]\right.\nonumber\\
                     & \qquad\qquad\qquad\qquad\left.{}+3[\gamma^2]^2-6[\gamma^4]
                       \right).
\label{eq:Pot}
\end{align}
\end{subequations}
Here $\lambda_A\,\,(A=0,1,2,3)$ are the eigenvalues of ${\gamma^\mu}_\nu$ and
we understand the square bracket notation as $[\gamma^k]\equiv tr(\gamma^k)$.

The variation of action (\ref{eq:action}) gives the bigravity field equations
\begin{equation}\label{eq:FieldEq}
{G^\mu}_\nu  =\frac{m^2\kappa_g}{\kappa}{V^\mu}_\nu, \qquad
{\mathcal{G}^\mu}_\nu =\frac{m^2\kappa_f}{\kappa}{\mathcal{V}^\mu}_\nu,
\end{equation}
where ${G^\mu}_\nu$ and ${\mathcal{G}^\mu}_\nu$ are the Einstein tensors for
$g_{\mu\nu}$ and $f_{\mu\nu}$, respectively, and the interaction
contributions are given by
\begin{subequations}\label{eq:Interactions}
\begin{align} {V^\mu}_\nu & \equiv
\frac{2g^{\mu\alpha}}{\sqrt{-g}}
\frac{\delta\left(\sqrt{-g}\mathcal{U}\right)}{\delta g^{\alpha\nu}}
={\tau^\mu}_\nu-\mathcal{U}{\delta^\mu}_\nu,\label{eq:Interaction_g}\\
{\mathcal{V}^\mu}_\nu & \equiv \frac{2f^{\mu\alpha}}{\sqrt{-f}}
\frac{\delta\left(\sqrt{-g}\mathcal{U}\right)}{\delta f^{\alpha\nu}}
=-\frac{\sqrt{-g}}{\sqrt{-f}}{\tau^\mu}_\nu,\label{eq:Interaction_f}
\end{align}
\end{subequations}
with
\begin{align}
{\tau^\mu}_\nu={}&
\left(b_1\mathcal{U}_0+b_2\mathcal{U}_1+b_3\mathcal{U}_2+b_4\mathcal{U}_3\right)
{\gamma^\mu}_\nu \nonumber \\
&-\left(b_2\mathcal{U}_0+b_3\mathcal{U}_1+b_4\mathcal{U}_2\right)
{(\gamma^2)^\mu}_\nu \nonumber \\
&+\left(b_3\mathcal{U}_0+b_4\mathcal{U}_1\right){(\gamma^3)^\mu}_\nu\nonumber\\
&-b_4\mathcal{U}_0{(\gamma^4)^\mu}_\nu.
\label{eq:EnergyMomentum2}
\end{align}

An alternative way to write the interaction potential is in terms of the
matrix ${\mathcal{K}^\mu}_\nu={\delta^\mu}_\nu-{\gamma^\mu}_\nu$, giving
\begin{equation}
\mathcal{U}[g,f]=\sum_{k=0}^{4}c_k\mathcal{U}_k(\mathcal{K}),
\label{eq:KapppaRep}
\end{equation}
where the interaction terms $\mathcal{U}_k(\mathcal{K})$ are again defined as
in (\ref{eq:Pot}) after the replacement $\gamma\rightarrow\mathcal{K}$, and
both sets of couplings constants $b_k$ and $c_k$ are linearly related. It is
desirable to exactly reproduce the Fierz-Pauli mass term in the weak field
limit, which is expressed by means of the matrix $\mathcal{K}$ as
\begin{equation}\label{eq:FP}
{}+\frac{m_{FP}^2}{2}\left(\left[\mathcal{K}\right]^2
-\left[\mathcal{K}^2\right]\right).
\end{equation}
Hence, it can be tracked down directly from the quadratic contribution of the
potential in the $\mathcal{K}$ formulation just by choosing $c_{2}=-1$ and
$\kappa=\kappa_{g}$, which makes the parameter $m^2$ in action
(\ref{eq:Bigravity}) become precisely the Fierz-Pauli mass in flat
spacetime. In other words, one of the coupling constants is not free since it
is chosen \emph{a priori } as the Fierz-Pauli mass. Returning to the $\gamma$
formulation this implies
\begin{equation}\label{eq:Cs2Bs}
-1=c_2=b_2+2b_3+b_4 \Rightarrow b_2=-1-2b_3-b_4.
\end{equation}
We will use this normalization of the couplings throughout our work, so the
constant $b_{2}$ will not appear as it is replaced in favor of $b_{3}$ and
$b_{4}$.

%%%%%%%%%%%%%%%%%%%%%%%%%%%%%%%%%%%%%%%%%%%%%%%%%%%%%%%%%%%%%%%%
\section{Kerr-Schild \emph{Ansatz} in Bigravity\label{Sec:Kerr-Schild}}
%%%%%%%%%%%%%%%%%%%%%%%%%%%%%%%%%%%%%%%%%%%%%%%%%%%%%%%%%%%%%%%%

The first nontrivial rotating solution in bigravity was found by Babichev and
Fabbri \cite{Babichev:2014tfa}; it consists of a pair of Kerr black holes
with different masses but rotating with the same angular momenta. In the
following section we present a rederivation of this solution following a
different approach. It is done to illustrate the procedure that will prove
useful later in deducing the rotating solutions with the cosmological constant.

The nonlinearity of Einstein field equations makes difficult any attempt to
find general solutions to a theory of gravity. However, there are many
strategies that can be followed to reduce the level of complexity, and one of
them is to start with an educated\emph{ Ansatz}. For instance, almost all known
black holes can be written as so-called Kerr-Schild transformations from the
spacetime defining their asymptotic behavior \cite{Schild:1965}. The
simplification of this \emph{Ansatz} in standard gravity lies in the fact that the
field equations become linearized exactly, i.e.,\ without approximations.

Because in bigravity we have two sets of Einstein field equations
(\ref{eq:FieldEq}) coupled by an interaction potential (\ref{eq:Int}), we can
expect a similar simplification for such kinds of \emph{Ans{\"a}tze}. Concretely, we
will assume the first metric as a Kerr-Schild transformation from Minkowski
spacetime and the second one as being proportional to a different Kerr-Schild
transformation also from Minkowski spacetime
\begin{subequations}\label{eq:Kerr-Schilds}
\begin{align}
ds_g^2=g_{\mu\nu}dx^\mu dx^\nu & =ds_M^2+2S_1l\otimes l,
\label{eq:gKerr-Schild}\\
ds_f^2=f_{\mu\nu}dx^\mu dx^\nu & =C^2\left(ds_M^2+2S_2l\otimes l \right),
\label{eq:fKerr-Schild}
\end{align}
\end{subequations}
where $ds_M^2$ is the Minkowski metric, $l$ is the tangent vector to a null,
geodesic, and shear-free congruence on Minkowski spacetime, $S_1$ and $S_2$ are
a pair of scalar profiles and $C$ is a dimensionless proportionality
constant. We are interested in describing stationary and axisymmetric
spacetimes with these \emph{Ans{\"a}tze}. Hence, the above ingredients must be
compatible with these symmetries and this is best realized in the so-called
ellipsoidal coordinates \cite{Gibbons:2004uw}, where the Minkowski
metric is written as
\begin{equation}\label{eq:MinkowskiET}
ds_M^2=-dt^2+\left(r^2+a^2\right)\sin^2\theta d\phi^2
+\frac{\Sigma}{r^2+a^2}dr^2+\Sigma d\theta^2,
\end{equation}
with $\Sigma=r^2+a^2\cos^2\theta$. These coordinates are understood if the
Cartesian spatial slices of Minkowski spacetime are foliated by ellipsoids of
revolution
\begin{equation}\label{eq:ellipsoids}
\frac{x^2+y^2}{r^2+a^2}+\frac{z^2}{r^2}=1,
\end{equation}
rather than standard spheres. Notice that the spheres are recovered for $a=0$;
consequently, the parameter $a$ denotes the departure from sphericity of the
ellipsoids. The coordinate $r$ labels each ellipsoid and the angular
coordinates $\theta$ and $\phi$ parametrize the ellipsoids as is easily
inferred from Eq.~(\ref{eq:ellipsoids}). The stationary and axisymmetric
isometries are represented in this coordinates by the Killing vectors
$k=\partial_t$ and $m=\partial_\phi$, respectively. The relevance of these
coordinates lies in the fact that it is possible to prove that in Minkowski
spacetime there exists only one congruence of shearfree null geodesics that
is at the same time stationary and axisymmetric \cite{Ayon-Beato:2015nvz}; these
coordinates define a parametrization where the related tangent vector can be
expressed in closed form as
\begin{equation}\label{eq:l}
l=dt-a\sin^2\theta d\phi+\frac{\Sigma}{r^2+a^2}dr.
\end{equation}
Finally, in order to respect the stationary and axisymmetric isometries the
profiles must be independent of the coordinates $t$ and $\phi$, i.e.\
$S_1=S_1(r,\theta)$ and $S_2=S_2(r,\theta)$. All these ingredients completely
determine a stationary and axisymmetric Kerr-Schild transformation from flat
spacetime.

Let us compute now the square of the matrix $\gamma$ according to its
definition (\ref{def:gamma}) by taking the product
\begin{align}
{(\gamma^2)^\mu}_\nu &= \left(\eta^{\mu\alpha}-2S_1l^{\mu}l^{\alpha}\right)
C^2\left(\eta_{\alpha\nu}+2S_2l_{\alpha}l_{\nu}\right)\nonumber\\
&= C^2\left[{\delta^\mu}_\nu-2(S_1-S_2)l^{\mu}l_{\nu}\right].
\label{eq:gamma2KerrSchild}
\end{align}
It is precisely here where the utility of the Kerr-Schild \emph{Ansatz} becomes
manifest, since, independently of the seed metric, the null character of the
involved vector field makes its contribution to the matrices nilpotent which
leads to a truncation of the matrix power expansion for the square root.
Hence, it is straightforward to find a closed form for the square root matrix
using the Kerr-Schild \emph{Ans{\"a}tze}
\begin{equation}\label{eq:gammaKerrSchild}
{\gamma^\mu}_\nu=C\left[{\delta^\mu}_\nu-(S_1-S_2)l^{\mu}l_{\nu}\right].
\end{equation}
Now we may proceed to write down the interaction terms
(\ref{eq:Interactions}), supported by the previously mentioned nilpotent property which
allows us to build any power of the square root matrix as
\begin{equation}
{(\gamma^n)^\mu}_\nu=C^n\left[{\delta^\mu}_\nu-n(S_1-S_2)l^{\mu}l_{\nu}\right],
\label{eq:gamma2n}
\end{equation}
arriving at these particularly simple expressions
\begin{subequations}\label{eq:InteractionTensors}
\begin{align}
{V^\mu}_\nu= & P_1{\delta^\mu}_\nu-CP_0(S_1-S_2)l^{\mu}l_\nu,
\label{eq:gInteractionTensors}\\
{\mathcal{V}^\mu}_\nu= & \frac{P_2}{C^3}{\delta^\mu}_\nu
+\frac{P_0}{C^3}(S_1-S_2)l^{\mu}l_\nu,
\label{eq:fInteractionTensors}
\end{align}
\end{subequations}
where the coefficients are linear combinations of the coupling constants:
\begin{subequations}\label{eq:Constraints}
\begin{align}
P_0&\equiv -2Cb_4+C(C-4)b_3+b_1-2C,\label{eq:P0}\\
P_1&\equiv 3C^2b_4-C^2(C-6)b_3-3Cb_1-b_0+3C^2,\label{eq:P1}\\
P_2&\equiv -C(C^2-3)b_4-3C(C-2)b_3-b_1+3C.\label{eq:P2}
\end{align}
\end{subequations}
In the following section, we use these expressions to prove a circularity
theorem, which is the basis of the integration procedure allowing us to obtain
rotating solutions from the Kerr-Schild \emph{Ansatz} \cite{Ayon-Beato:2015nvz}.

%%%%%%%%%%%%%%%%%%%%%%%%%%%%%%%%%%%%%%%%%%%%%%%%%%%%%%%%%%%%%%
\section{A circularity theorem\label{sec:Circularity}}
%%%%%%%%%%%%%%%%%%%%%%%%%%%%%%%%%%%%%%%%%%%%%%%%%%%%%%%%%%%%%%

We start by recalling that for any stationary axisymmetric spacetime with
commuting Killing vector fields $k=\partial_t$ and $m=\partial_\phi$, the
following geometrical identities hold \cite{Heusler:1996}:
\begin{subequations}\label{eq:EEprojected}
\begin{align}
C_{k}& \equiv  d*\left(k\wedge m\wedge dk\right)
-2*\left(k\wedge m\wedge R(k)\right)=0,\label{eq:EE1}\\
C_{m}& \equiv d*\left(k\wedge m\wedge dm\right)
-2*\left(k\wedge m\wedge R(m)\right)=0.\label{eq:EE2}
\end{align}
\end{subequations}
Here the Killing fields are understood as one-forms, $k=g_{\mu\nu}k^\nu
dx^\mu$ and $m=g_{\mu\nu}m^\nu dx^\nu$, while the Ricci one-forms amount to
$R(k)=R_{\mu\nu}k^\nu dx^\mu$ and $R(m)=R_{\mu\nu}m^\nu dx^\mu$. These
identities are the basis of the so-called circularity theorem  in general
relativity; in vacuum they imply that the functions under the differential are
constants, which in turn must vanish at the symmetry axis where $m=0$;
consequently,
\begin{equation}\label{eq:Frobenius}
k\wedge m\wedge dk=0=k\wedge m\wedge dm.
\end{equation}
These are the Frobenius integrability conditions defining circularity; i.e.\,
the planes orthogonal to the Killing vectors at any point are integrable to
surfaces orthogonal to the Killing fields in the whole spacetime. Choosing
coordinates along Killing fields and on their orthogonal surfaces the
circular metric becomes block diagonal; the iconic example are the well-known
Boyer-Lindquist coordinates \cite{Boyer:1966qh} of the Kerr black hole
\cite{Kerr:1963ud}.

The standard circularity argument of general relativity for stationary and
axisymmetric configurations cannot be straightforwardly extended to bigravity
due to the nontrivial interaction terms (\ref{eq:Interactions}). However, for
the Kerr-Schild \emph{Ans{\"a}tze} (\ref{eq:Kerr-Schilds}), which are not circular
by construction, we will establish a circularity theorem. This result will
fix the angular dependencies of the involved profiles imposing at the same
time a constraint between the coupling constants. This theorem reduces each
Einstein equation to a single independent equation that we easily integrate
in the rest of the section.

We start by calculating the following quantities, using the Kerr-Schild
\emph{Ansatz} (\ref{eq:gKerr-Schild}) in both the definition of the Killing
one-forms and the interaction term (\ref{eq:gInteractionTensors}),
\begin{equation}\label{eq:kmFormsRelation}
\frac{*\left(k\wedge m\wedge
dk\right)}{l_t}=\frac{*\left(k\wedge m\wedge dm\right)}{l_\phi}
=-\frac{2l_{t}\sin\theta}{\Sigma}\partial_{\theta}(\Sigma
S_1),
\end{equation}
\begin{align}
\frac{*\left(k\wedge m\wedge V(k)\right)}{l_t}&=
\frac{*\left(k\wedge m\wedge V(m)\right)}{l_\phi}\nonumber\\
&=-CP_0(S_1-S_2)*(k\wedge m\wedge l),
\label{eq:VFormsRelation}
\end{align}
where the interaction one-forms are defined analogously to the Ricci
one-forms as $V(k)=V_{\mu\nu}k^\nu dx^\mu$ and $V(m)=V_{\mu\nu}m^\nu dx^\mu$.
Using Einstein equations for the metric $g_{\mu\nu}$ in the identities
(\ref{eq:EEprojected}) and taking into account the explicit expressions of
the above quantities, we arrive at the following identity
\begin{equation}\label{eq:Circularity}
\frac{l_\phi}{l_t}C_{k}-C_{m}=
-*(k\wedge m\wedge dk) d\left(\frac{l_\phi}{l_t}\right)=0,
\end{equation}
which implies the circularity conditions (\ref{eq:Frobenius}). Using the
explicit expressions (\ref{eq:kmFormsRelation}), the circularity
automatically fix the angular dependence of the profile. Additionally, taking
into account the circularity in the identities (\ref{eq:EEprojected})
together with Einstein equations necessarily implies that the expressions
(\ref{eq:VFormsRelation}) must also be identically zero. This imposes a
constraint between the coupling constants for the nontrivial case of the
different profiles. Exactly the same can be concluded for the second metric;
we can repeat the same arguments for the  equations anologous to
(\ref{eq:EEprojected})-(\ref{eq:Circularity}) built from $f_{\mu\nu}$. Hence,
we establish a circularity theorem for both metrics which has as its consequences
\begin{equation} \label{eq:flatSi}
\displaystyle{S_{i}(r,\theta)=\frac{rM_{i}(r)}{\Sigma}}, \quad P_{0}=0,
\quad i=1,2.
\end{equation}

Now the integration of the remaining Einstein equations is straightforward.
Bearing in mind the circularity restrictions, the interaction terms
(\ref{eq:InteractionTensors}) reduce to the diagonal form
\begin{equation}
{V^\mu}_\nu=P_1{\delta^\mu}_\nu,\hspace{5mm}
{\mathcal{V}^\mu}_\nu=\frac{P_2}{C^3}{\delta^\mu}_\nu,
 \label{eq:V_eval}
\end{equation}
and the only independent equations for each Einstein set are the following
combinations
\begin{subequations}\label{eq:EEcomb}
\begin{align}
\frac{2rM_1}{\Delta_1}\!\left({G^r}_t+\frac{a}{r^2+a^2}{G^r}_\phi\right)&
\nonumber\\
-\left({G^r}_r-\frac{m^2\kappa_g}{\kappa}{V^r}_r\right)&=
\frac{2r^2}{\Sigma^2}M_1'+\frac{m^2\kappa_g}{\kappa}P_{1}=0,
  \label{eq:EEcomb_g}\\
\frac{2C^{2}rM_{2}}{\Delta_{2}}\!\left({\mathcal{G}^r}_t
+\frac{a}{r^2+a^2}{\mathcal{G}^r}_\phi\right)&\nonumber\\
{}-C^{2}\!\left({\mathcal{G}^r }_r
-\frac{m^2\kappa_f}{\kappa}{\mathcal{V}^r}_r\right)&=
\frac{2r^2}{\Sigma^2}M_2'+\frac{m^2\kappa_f}{\kappa}\frac{P_{2}}{C}=0,
\label{eq:EEcomb_f}
\end{align}
\end{subequations}
with $\Delta_{i}=r^2+a^2-2rM_{i}(r)$, $i=1,2$. Because $\Sigma$ carries the
only $\theta$-dependence on the right-hand side, the only way to fulfill
these equations is if each term independently vanishes, implying
$M_1(r)=m_1$, $M_2(r)=m_2$, and $P_1=0=P_2$, with $m_1$ and $m_2$ being
independent integration constants. The rest of the Einstein equations are
automatically satisfied. Finally, \emph{the most general family of stationary
axisymmetric Kerr-Schild transformations from flat spacetime solving
bigravity equations is}
\begin{subequations}\label{eq:KerrSol}
\begin{align}
ds_g^2 & =ds_M^2+\frac{2m_{1}r}{\Sigma}l\otimes l ,
\label{eq:gKerrSol}\\
ds_f^2 & =C^2\left(ds_M^2+\frac{2m_{2}r}{\Sigma}l\otimes l \right),
\label{eq:fKerrSol}\\
P_{0} & = P_1 = P_2 = 0,\label{eq:constrsKerrSol}
\end{align}
\end{subequations}
where the Minkowski metric and the null vector $l$ written in ellipsoidal
coordinates are given by (\ref{eq:MinkowskiET}) and (\ref{eq:l}),
respectively. Additionally, the constraints between the coupling constants
(\ref{eq:constrsKerrSol}) are read from definitions (\ref{eq:Constraints})
and are not satisfied for $C=1$, which justify the use of the proportionality
constant. This solution corresponds to a pair of Kerr black holes
\cite{Kerr:1963ud} with the same angular momenta but different masses. It was
originally obtained in \cite{Babichev:2014tfa} by direct substitution. The
link between the coupling constants used in their work and ours is the
following
\begin{subequations}
\begin{align}
b_{0}&=-\frac{3\alpha+\beta-\Lambda_{g}+3}{1-\bar{\kappa}\Lambda_{f}}, \,
b_{1}=\frac{2\alpha+\beta+1}{1-\bar{\kappa}\Lambda_{f}}, \,
b_{3}=\frac{\beta}{1-\bar{\kappa}\Lambda_{f}}, \,
\nonumber\\
b_{4}&=\frac{\alpha-\beta+\bar{\kappa}\Lambda_{f}-1}
{1-\bar{\kappa}\Lambda_{f}}, \,
\frac{\kappa_{g}}{\kappa}m^2=(1-\bar{\kappa}\Lambda_{f})\bar{m}^2,
\end{align}
\end{subequations}
where in Ref.~\cite{Babichev:2014tfa}, $\bar{m}$ is the mass,
$\bar{\kappa}$ is the ratio of Einstein constants for both metrics, and
$\Lambda_{g,f}$ are the dimensionless cosmological constants.

It is worth noting that although the circularity theorem applies to both
metrics, one could find the Boyer-Lindquist coordinates to block-diagonalize
one of them, but since the masses are not equal, those coordinates are not
suitable to diagonalize simultaneously the second metric, a fact already
noticed by Babichev and Fabbri \cite{Babichev:2014tfa}.

%%%%%%%%%%%%%%%%%%%%%%%%%%%%%%%%%%%%%%%%%%%%%%%%%%%%%%%%%%%%%%%%%%%%%%%
\section{Kerr-(A)dS black holes in bigravity\label{Sec:Kerr-deSitter}}
%%%%%%%%%%%%%%%%%%%%%%%%%%%%%%%%%%%%%%%%%%%%%%%%%%%%%%%%%%%%%%%%%%%%%%%

The rotating solutions found by Babichev and Fabbri in
\cite{Babichev:2014tfa} are without doubt an interesting and nontrivial
result extending the scope of the physics of black holes that can be
understood under a ghost-free dynamics. At the same time, it is a little
surprising to find just asymptotically flat configurations since the coupling
constants $b_0$ and $b_4$ in action (\ref{eq:action}) play the role of
cosmological constants for each metric. This suggests the possibility that
rotating configurations can be generalized to include asymptotic behaviors
with nontrivial constant curvature. These spacetimes are very well-known in
general relativity in the presence of a cosmological constant and correspond to
the Kerr-(A)dS black hole originally discovered by Carter in
\cite{Carter:1968ks}. The possibility of their inclusion within the bigravity
vacua was also discussed in Ref.~\cite{Babichev:2014tfa}; they use rotating
generalizations of the Eddington-Finkelstein null coordinates that they
intend to generalize in the presence of a cosmological constant. As we identify
in the previous section, the success in absence of a cosmological constant is
due more to the underling Kerr-Schild structure. Fortunately, Carter himself
presented the Kerr-(A)dS black holes in a generalized Kerr-Schild form
starting from the (anti-)de Sitter spacetime \cite{Carter:1973rla}, this form
has even been amenable to higher-dimensional extensions
\cite{Gibbons:2004uw}. This will be precisely our starting point in the
search for a generalization of the solutions \cite{Babichev:2014tfa}. We use
the fact that the Kerr-Schild \emph{Ans{\"a}tze} (\ref{eq:Kerr-Schilds}) can be
generalized by taking as seed the (A)dS spacetime instead of the flat one
\begin{subequations}\label{eq:Kerr-SchildsAdS}
\begin{align}
ds_g^2=g_{\mu\nu}dx^\mu dx^\nu & =ds_{0}^2+2S_1l\otimes l ,
\label{eq:gKerr-SchildAdS}\\
ds_f^2=f_{\mu\nu}dx^\mu dx^\nu & =C^2\left(ds_{0}^2+2S_2l\otimes l \right),
\label{eq:fKerr-SchildAdS}
\end{align}
\end{subequations}
where, in the conventions of \cite{Gibbons:2004uw}, the (A)dS metric in
ellipsoidal coordinates is
\begin{align}
ds_0^2= & -\frac{(1-\lambda r^2)\Delta_{\theta}}{1+\lambda a^2}dt^2
+\frac{(r^2+a^2)\sin^2{\theta}}{1+\lambda a^2}d\phi^2\nonumber\\
& +\frac{\Sigma}{(1-\lambda r^2)(r^2+a^2)}dr^2
+\frac{\Sigma}{\Delta_{\theta}}d\theta^2.
\label{eq:(A)dSMetric}
\end{align}
Here, $\Delta_\theta=1+\lambda a^2\cos^2\theta$, again
$\Sigma=r^2+a^2\cos^2\theta$ and the constant curvature is fixed by the
scalar $R=12\lambda$, which determines the effective cosmological constant
$\Lambda_{\mathrm{eff}}=3\lambda$, being $\lambda=\pm1/\ell^2$ the inverse of
the square (A)dS radius. The null, geodesic and shearfree vector field on
(A)dS is \cite{Gibbons:2004uw}
\begin{equation}\label{eq:ldS}
l=\frac{\Delta_\theta}{1+\lambda a^2}dt
-\frac{a\sin^2\theta}{1+\lambda a^2}d\phi
+\frac{\Sigma}{(1-\lambda r^2)(r^2+a^2)}dr.
\end{equation}
In this coordinates again the stationary and axisymmetric isometries are
manifest if the scalar profiles are also invariant along the Killing fields
$k=\partial_t$ and $m=\partial_\phi$, by choosing $S_1=S_1(r,\theta)$ and
$S_2=S_2(r,\theta)$. The Minkowski limit of Sec.~\ref{Sec:Kerr-Schild} is
consistently recovered for vanishing curvature, $\lambda=0$.

The key to success in the Kerr-Schild transformations is that the form of the
square root matrix (\ref{eq:gammaKerrSchild}) is unaltered by the replacement
in the ansatz described before; this implies the interaction terms remain the
same as in (\ref{eq:InteractionTensors}), so the procedure will closely
resemble that one where the Minkowski spacetime is chosen as the seed. In
fact, the circularity theorem of Sec.~\ref{sec:Circularity} applies unchanged
since Eqs.~(\ref{eq:kmFormsRelation})-(\ref{eq:Circularity}) looks exactly
the same for $\lambda\neq0$. As consequence, one arrives again to the
circular profiles and the same constraint (\ref{eq:flatSi}). As we have seen,
the cicularity implies the diagonalization of the interaction terms
(\ref{eq:InteractionTensors}), and the fact that there is only one
independent equation for each Einstein system
\begin{subequations}\label{eq:EEcomb_dS}
\begin{align}
&\frac{2rM_1}{\Delta^{r}_{1}}\left(\frac1{1-\lambda r^2}{G^r}_t
+\frac{a}{r^2+a^2}{G^r}_\phi\right)\nonumber\\
&-  \left({G^r }_r-\frac{m^2\kappa_g}{\kappa}{V^r}_r\right)=
\frac{2r^{2}M_{1}'}{\Sigma^2}+\frac{m^2\kappa_gP_{1}}{\kappa}+3\lambda=0,
\label{eq:EEcomb_gdS}\\
& \frac{2C^{2}rM_{2}}{\Delta^{r}_{2}}\left(
\frac1{1-\lambda r^2}{\mathcal{G}^r}_t
+\frac{a}{r^2+a^2}{\mathcal{G}^r}_\phi\right)
\nonumber\\
&-C^{2}\!\left({\mathcal{G}^r}_r
-\frac{m^2\kappa_f}{\kappa}{\mathcal{V}^r}_r\right)=
\frac{2r^{2}M_2'}{\Sigma^2}+\frac{m^2\kappa_fP_2}{\kappa C}+3\lambda=0,
\label{eq:EEcomb_fdS}
\end{align}
\end{subequations}
where $\Delta^{r}_{i}=r^2+a^2-2rM_{i}(r)-\lambda r^2(r^2+a^2)$. Again the
dependence in $\theta$ is fixed and the above equations are only satisfied if
each functionally independent term on $\theta$ vanishes separately. Which
means $M_1(r)=m_1$ and $M_2(r)=m_2$, with $m_1$ and $m_2$ arbitrary and
independent integration constants, and new constraints on the coupling
constants. Hence, we can write the new solution as
\begin{subequations}\label{eq:Kerr-dSSol}
\begin{align}
ds_g^2 &=ds_0^2+\frac{2m_1r}{\Sigma}l\otimes l,\label{eq:gKerr-dSSol} \\
ds_f^2 &=C^2\left(ds_0^2+\frac{2m_2r}{\Sigma}l\otimes l\right),
\label{eq:fKerr-dSSol}\\
P_{0}& =0,\quad \kappa_{g}P_{1}=\frac{\kappa_{f}P_{2}}{C}=
-\frac{3\lambda\kappa}{m^{2}},
\label{eq:constrsKerr-dSSol}
\end{align}
\end{subequations}
where the (A)dS metric, $ds_0^2$, in ellipsoidal coordinates is given in
(\ref{eq:(A)dSMetric}), their null vector $l$ is (\ref{eq:ldS}) and
definitions (\ref{eq:Constraints}) determine the new constraints
(\ref{eq:constrsKerr-dSSol}) for the coupling constants allowing the single
effective cosmological constant $\Lambda_{\mathrm{eff}}=3\lambda$. This
solution corresponds to \emph{a family of stationary-axisymmetric Kerr-Schild
transformations from (anti-)de Sitter spacetime} which describes two
Kerr-(A)dS black holes with the same angular momenta and (A)dS radii but
possessing independently defined masses.

Now it is possible to have a trivial proportionality constant $C=1$, but at
the cost of constraining the effective cosmological constant as
$\Lambda_{\mathrm{eff}}=3\lambda=-(\kappa_{f}/\kappa)m^2$ (notice that $m^2$
is no longer the mass square in this context, hence, it is just a no
necessarily positive coupling constant). Additionally, just as in the
asymptotically flat problem, both metrics are circular but it is not possible
to diagonalize them together through the same Boyer-Lindquist-like
transformation.

%%%%%%%%%%%%%%%%%%%%%%%%%%%%%%%%%%%%%%%%%%%%%%%%%%%%%%%%%%%%%
\section{Discussion \label{sec:discussion}}
%%%%%%%%%%%%%%%%%%%%%%%%%%%%%%%%%%%%%%%%%%%%%%%%%%%%%%%%%%%%%

In this paper we have explored the consequences of using Kerr-Schild
transformations for the dynamics of ghost-free bigravity. The first case
studied corresponded to rotating asymptotically flat spacetimes producing
the solution of two Kerr black holes with different masses already reported
by Babichev and Fabbri \cite{Babichev:2014tfa}. We managed to advance a
little further in the comprehension of how this configuration appears since
our \emph{Ans{\"a}tze} were not the Kerr black holes themselves but rather the most
general stationary-axisymmetric Kerr-Schild transformations from flat
spacetime \cite{Ayon-Beato:2015nvz}. It is particulary simple to calculate the
interaction terms in this case; the null character of the Kerr-Schild vector
gives rise to a nilpotent contribution; consequently, any matrix power series
is necessarily truncated and in particular, especially the one defining the square root
matrix encoding the interactions. This allows us to find that the dynamics of
the theory imposes a circularity theorem applicable to both metrics. As a
consequence, the resulting profiles necessarily have to be those of the Kerr
spacetime, but for independent masses. No other rotating solution of massive
(bi)gravity with a flat asymptotic can be constructed in this manner.

Of course, the existence of more general rotating solutions, whose
integration is tackled by different procedures, still remains an open
problem. For example, an interesting question is if it is possible to have
solutions not only with different masses but also with different angular
momenta. This question was already posed in Ref.~\cite{Babichev:2014tfa},
where no definite answer was given due to the manifest difficulty of the
involved calculations. A starting point to explore this question is to
consider the slight modification of the Kerr-Schild \emph{Ans{\"a}tze}
(\ref{eq:Kerr-Schilds}), in which the seed flat spacetimes of each
metric are foliated by revolution ellipsoids (\ref{eq:ellipsoids}) with
different ellipticity parameters $a_1$ and $a_2$. The shear-free null
geodesics of each differently foliated flat spacetime will be accordingly
parameterized and unfortunately the nilpotent property giving rise to the
simple expression (\ref{eq:gammaKerrSchild}) for the square root matrix will no longer apply. However, in this case the square root matrix can still be
calculated by going to the tetrad formalism, where the chosen tetrads must
satisfy the so-called symmetrization condition which warrants their
equivalence to the metric formulation\cite{Hinterbichler:2012cn,Volkov:2012zb}. It is possible to
show that the symmetrization condition necessarily requires that $a_1=a_2$,
which returns us to the setting analyzed in our paper. This points to a
possible obstruction to the existence of two ghost-free interacting metrics
with different angular momenta. Interestingly, this would have the advantage
that both metrics become singular exactly at the same spacetime locus: the
famous ringlike singularity of the Kerr black hole [corresponding to
$\Sigma=0$, or, equivalently, the interception of the plane $z=0$ with the
$r=0$ revolution ellipsoid (\ref{eq:ellipsoids})]. In other words, two
different angular momenta would suppose the existence of two ringlike
singularities making even more complex the causal structure of a rotating
bigravity spacetime.

Another generalization also discussed in Ref.~\cite{Babichev:2014tfa} is the
possibility of having asymptotically (A)dS rotating spacetimes. Their strategy,
based on the use of rotating generalizations of the Eddington-Finkelstein
null coordinates, ends up being too difficult to apply in the presence of a
cosmological constant. However, once one identifies that the fundamental
underlying null structure is the one associated with a Kerr-Schild
transformation, the generalization to include an asymptotic cosmological
constant is straightforward. This is the focus of the second case under
study, where we manage to repeat the arguments and find a new rotating
solution to the bigravity equations consisting of two Kerr-(A)dS black holes
sharing the same angular momentum and (A)dS radius but having independent
masses. Three constraints between the coupling constants and the
proportionality constant defining the second metric are needed in order to
allow this configuration. The first one is a result of the circularity of
these backgrounds and is independent of the existence of an effective
cosmological constant. The other two constraints define the same effective
cosmological constant for each metric. Thanks to the nontrivial character of
this effective cosmological constant, the bigravity interaction potential is
no longer vanishing, which means these black holes are not decoupled as in
the asymptotically flat case. Seemingly, one could think of generalizing this
result by considering two different (A)dS radii $\lambda_{1}$ and
$\lambda_{2}$ for each metric in \emph{Ans{\"a}tze} (\ref{eq:Kerr-SchildsAdS}). This
consideration would spoil, of course, the simplifications in finding the square
root matrix brought by the Kerr-Schild \emph{Ansatz}, but we could switch to the
tetrad formalism and go forward. What we found is that even letting the
angular momenta and cosmological constants be unconstrained, the same
symmetrization condition as in the asymptotically flat case requires both
angular momenta and both cosmological constants to be equal, making our
consideration in this sense the most general.

Finally, with our approach it is very easy to charge one of the two
asymptotically (A)dS black holes by coupling the Maxwell field to the
corresponding metric, let us say $g_{\mu\nu}$. This was first shown in the
asymptotically flat case by Babichev and Fabbri \cite{Babichev:2014tfa}. The
Kerr-Schild \emph{Ansatz} is extended to the charged case by choosing the vector
potential as proportional to the null vector $l$, which in (A)dS is
(\ref{eq:ldS}). It is an easy task to solve Maxwell equations guided by the
circularity that must satisfy a stationary axisymmetric electromagnetic field
\cite{Heusler:1996}; the result is $A=qrl/\Sigma$ where $q$ is the electric
charge. Consequently, the profiles are again the circular ones and only the
first of them becomes modified since now $M_1(r)=m_1-q^2/2r$, giving a
Kerr-Newmann-(A)dS black hole coupled to a Kerr-(A)dS one with the same
properties and constraints than before. One can also ask what kind of
rotating configurations could lead to coupling the Maxwell field nonminimally
via the effective metric that has been intensely analyzed recently in the
literature \cite{deRham:2014naa}. It is possible to show that our approach is
incompatible with this nonminimal coupling and, therefore, to favor only the minimal coupling to a single metric, as we just described.

\begin{acknowledgments}
This work has been funded by grants 175993, 178346, 243342 and 243377 from
CONACyT, together with grants 1121031, 1130423 and 1141073 from FONDECYT. EAB
was partially supported by the ``Programa Atracci\'{o}n de Capital Humano
Avanzado del Extranjero, MEC'' from CONICYT. DHB and JAMZ were supported by
the ``Programa de Becas Mixtas'' from CONACyT.
\end{acknowledgments}

\end{document}